\documentclass[aps,prd,amssymb,showpacs,floatfix,preprint,nofootinbib,superscriptaddress]{revtex4-1}
\usepackage{amsmath,amsfonts,graphics,epsfig,amssymb,color}

\definecolor{DarkGreen}{RGB}{0,0,127}

\newcommand{\red}{\protect\color{red}}
\newcommand{\blue}{\protect\color{blue}}
\newcommand{\black}{\protect\color{black}}

\renewcommand\blue\red
\renewcommand\red\black

\begin{document}
\title{Constraints on the flux of $\sim (10^{16} - 10^{17.5})$~eV cosmic
photons from the EAS--MSU muon data}

\author{Yu.\,A.\,Fomin}
\affiliation{D.V.~Skobeltsyn Institute of Nuclear Physics,
M.V.~Lomonosov Moscow State University, Moscow 119991, Russia}
\author{N.\,N.\,Kalmykov}
\affiliation{D.V.~Skobeltsyn Institute of Nuclear Physics,
M.V.~Lomonosov Moscow State University, Moscow 119991, Russia}
\author{I.\,S.\,Karpikov}
\affiliation{Institute for Nuclear
Research of the Russian Academy of Sciences, 60th October Anniversary
Prospect 7a, Moscow 117312, Russia}
\author{G.\,V.\,Kulikov}
\affiliation{D.V.~Skobeltsyn Institute of Nuclear Physics,
M.V.~Lomonosov Moscow State University, Moscow 119991, Russia}
\author{M.\,Yu.\,Kuznetsov}
\affiliation{Institute for Nuclear
Research of the Russian Academy of Sciences, 60th October Anniversary
Prospect 7a, Moscow 117312, Russia}
\author{G.\,I.\,Rubtsov}
\affiliation{Institute for Nuclear
Research of the Russian Academy of Sciences, 60th October Anniversary
Prospect 7a, Moscow 117312, Russia}
\author{V.\,P.\,Sulakov}
\affiliation{D.V.~Skobeltsyn Institute of Nuclear Physics,
M.V.~Lomonosov Moscow State University, Moscow 119991, Russia}
\author{S.\,V.\,Troitsky}
\email{st@ms2.inr.ac.ru}
\affiliation{Institute for Nuclear
Research of the Russian Academy of Sciences, 60th October Anniversary
Prospect 7a, Moscow 117312, Russia}



\begin{center}
\begin{abstract}
Results of the search for $\sim (10^{16} - 10^{17.5})$~eV primary
cosmic-ray photons with the data of the Moscow State University (MSU)
Extensive Air Shower (EAS) array are reported. The full-scale reanalysis
of the data with modern simulations of the installation does not confirm
previous indications of the excess of gamma-ray candidate events. Upper
limits on the corresponding gamma-ray flux are presented. The limits are
the most stringent published ones at energies $\sim 10^{17}$~eV.
\end{abstract}
\end{center}
\maketitle

\section{Introduction}
\label{sec:intro}
Searches for astrophysical gamma rays with energies $10^{15}$~eV$\lesssim
E_{\gamma} \lesssim 10^{20}$~eV attracted considerable attention for
years~\cite{1975, Risse}. At these energies, photons interact with the
atmosphere and produce extensive air showers (EAS) which may be detected
by installations studying cosmic rays. The primary motivation for the
studies includes multimessenger astronomy. Photons work as a diagnostic
tool to distinguish between various models of the origin of energetic
cosmic rays and neutrinos. In particular, because of the pair production on
cosmic background radiation, the photon flux from extragalactic sources is
strongly suppressed at sub-PeV to sub-EeV energies, and non-observation of
gamma rays in this energy band would strongly favor the extragalactic
origin of IceCube-detected astrophysical neutrinos over the Galactic
models \cite{Ahlers, Kalashev-ST-gamma}. At higher energies,
non-observation of cosmogenic photons \cite{BerZats} would strongly
constrain models with proton composition of cosmic rays with $E \gtrsim
10^{19.5}$~eV, see e.g.\ Refs.~\cite{Gelmini, Sarkar}. Contrary,
observation
of high-energy photons may be a smoking-gun signal of new physics,
including superheavy dark matter \cite{SHDM, SHDM1} (see also
Ref.~\cite{KRT-SHDM} and, for a recent reanalysis,
Ref.~\cite{Kuznetsov}), axion-like particles \cite{axion}, or
an ultimate test of Lorentz-invariance \cite{Lorentz-violation, GrishaLIV}.

For air-shower experiments, the main problem in the photon search is to
separate gamma-ray induced events from the usual, hadron-induced air
showers. One of the best discriminating variables is the muon number of a
EAS, which is much lower in photon-induced events, compared to the bulk of
showers. Indeed, a photon-induced shower develops by means of
electromagnetic interactions mostly, and the only source of muons is
provided by photonuclear reactions, which have a relatively small cross
section. Contrary, in a hadron-induced shower, lots of muons are produced
in decays of $\pi$ mesons, born in hadronic interactions. Unfortunately,
muon detectors are small or missing in many modern cosmic-ray experiments.
Here, we take advantage of having a large muon detector in the EAS-MSU
experiment and use its data to search for air showers poor in muons.

Previous preliminary studies indicated some excess of gamma-ray candidates
in the EAS-MSU data with respect to the expected background~\cite{Khorkhe,
  EAS-MSU_gamma_2013, EAS-MSU_gamma_2014}. These results invited for a
detailed study of the data with state of art methods. This study has been
performed: a full Monte-Carlo model of the data and the detector was
constructed~\cite{EAS-MSU_Full_MC} which gives an excellent description of
the surface-detector results~\cite{EAS-MSU_Full_MC} as well as of the
bulk of muon-detector data~\cite{EAS-MSU_mu}. Here, we present the
ultimate results of this study and consider muonless photon-candidate
events. We will see that, within the present approach, no excess of these
photon-like events is seen in the data. This allows us to put upper limits
on the primary gamma-ray flux which are world-best at energies $E_{\gamma}
\sim 10^{17}$~eV.

The rest of the paper is organized as follows. In Sec.~\ref{sec:data}, we
briefly describe the installation, the data and the Monte-Carlo
simulations used in this work; references to more detailed descriptions
are given. We derive the main result of the paper, the limits on the
gamma-ray flux, in Sec.~\ref{sec:flux}. In Sec.~\ref{sec:disc}, we
estimate systematic uncertainties of the result and compare it with other
studies, including previous photon searches in the EAS-MSU experiment. We
briefly conclude in Sec.~\ref{sec:concl}.

\section{Data and simulations}
\label{sec:data}
The EAS-MSU experiment and the Monte-Carlo model are described in
detail in the previous papers \cite{EAS-MSU_Full_MC, EAS-MSU}.
The installation consists of 76 charged-particle detector stations, in
which multiple Geiger-Mueller counters are located which determine the
number of charged particles ($N_{e}$) in a shower. The total number of
Geiger-Mueller counters in the surface part of the installation is
about 10,000. The total area of all Geiger-Mueller counters is $\sim
250$~m$^{2}$; the total area of the installation is $\sim
0.5$~km$^{2}$. In this work, we use the data of the large muon
detector located at the center of the array at a depth of 40 meters of
water equivalent underground corresponding to the threshold energy of
10~GeV for vertical muons. The muon detector had the total area of
36.4~m$^2$ and consisted of Geiger-Mueller counters with area of
0.033~m$^{2}$ each.

For the present study, we select events with the following cuts:
\begin{itemize}
 \item
the event passes the reconstruction procedure and the reconstruction
quality criteria are satisfied \cite{book};
 \item
the age parameter of the EAS is in the range $0.3<S<1.8$;
 \item
the reconstructed zenith angle $\theta$ is below $30^{\circ}$;
\item
the distance between the reconstructed shower axis and the array center
where the muon detector is located is $R<240$~m;
\item
the reconstructed shower size is $N_{e}>10^{7}$;
\item
out of 32 sections of the muon detector, at least 28 were operational.
\end{itemize}
The only difference with the cuts used in Ref.~\cite{EAS-MSU_mu} is
the lower $N_{e}$ threshold which is chosen to cover a wider range of
energies of potential photons. The extension of the $N_{e}$ range
  requires a new comparison of data to Monte-Carlo to validate the
  simulations. The corresponding distributions of $N_{e}$, shower age
  $S$, core distance of the array center $R$, zenith angle $\theta$
  and muon density at 100 meters $\rho_{\mu}(100)$ are shown in
  Appendix.  With the account of the muon detector operation cut, we
are left with 1204 days of data taking within 1984--1990. The data set
contains 3148 events.

Following the previous studies, we consider the following criteria to
select photon-candidate events: the muon detector
is not triggered by the shower.
However, low-energy protons do not always produce a sufficient number of
muons in EAS to activate the muon detector, in particular, at large
distances between the detector and the shower axis.
To evaluate the background of muonless events from hadronic primaries, we
make use of the full Monte-Carlo (MC) model of the EAS-MSU array described
in Ref.~\cite{EAS-MSU_Full_MC}.
It includes simulations of artificial air showers by the CORSIKA
 7.4001 \cite{Heck:1998vt} package
with the  QGSJET-II-04 \cite{Ostapchenko:2010vb} high-energy hadronic
 interaction model,
FLUKA 2011.2c \cite{Fluka} low-energy hadronic interaction model
and EGS4 \cite{EGS4} electromagnetic model.
Artificial EAS are recorded and processed identically to the experimental
data.
The Monte-Carlo simulations have been performed for proton and iron
primaries, see Ref.~\cite{EAS-MSU_Full_MC} for details; a realistic
composition was assumed to be a mixture of the two. The muon component of
EAS is highly dependent on the primary composition, therefore the number
of muonless background events depends on the assumed proton fraction.
For this study, the primary composition has been
determined  in Ref.~\cite{EAS-MSU_mu} by fitting the observed distribution
of the muon densities.
The assumed fraction of protons is $46 \pm 6 \%$. We note that, as discussed in
Ref.~\cite{EAS-MSU_mu}, it agrees well with the composition obtained from
the EAS-MSU surface-detector data~\cite{EAS-MSU_Full_MC}. More details on
the MC set with hadronic primaries may be found in
Ref.~\cite{EAS-MSU_Full_MC}.

We also need a MC simulation with gamma-ray primaries which is used to
determine the efficiency of the installation for the gamma-ray detection,
to relate the reconstructed $N_{e}$ to the primary-photon energy
$E_{\gamma}$ and to account for the amount of photon showers which do not
produce photon-candidate (that is, muonless) events. The simulation and
reconstruction of the artificial photon showers was performed in a way
similar to that for hadron-induced events~\cite{EAS-MSU_Full_MC}. The
total number of simulated independent gamma-induced showers is 300, their
thrown energies follow the $E_{\gamma}^{-1}$ spectrum with
  lower bound of MC $10^{16}<E_{\gamma} \le
10^{17.5}$~eV, and they are selected at the reconstruction stage to
reproduce the primary spectrum $\sim E_{\gamma}^{-2}$, as is customary in
high-energy photon searches. The dependence of the resulting limits on the
assumed spectrum is through the efficiency only and is weak. The
total number of MC realizations of gamma-induced events is 27310 (see
Ref.~\cite{EAS-MSU_Full_MC} for description of the sampling). Of them,
3898 passed all cuts.

\section{The gamma-ray flux limit}
\label{sec:flux}
\begin{figure}

\begin{minipage}[h]{0.48\linewidth}
\center{
\includegraphics[width=1\linewidth]{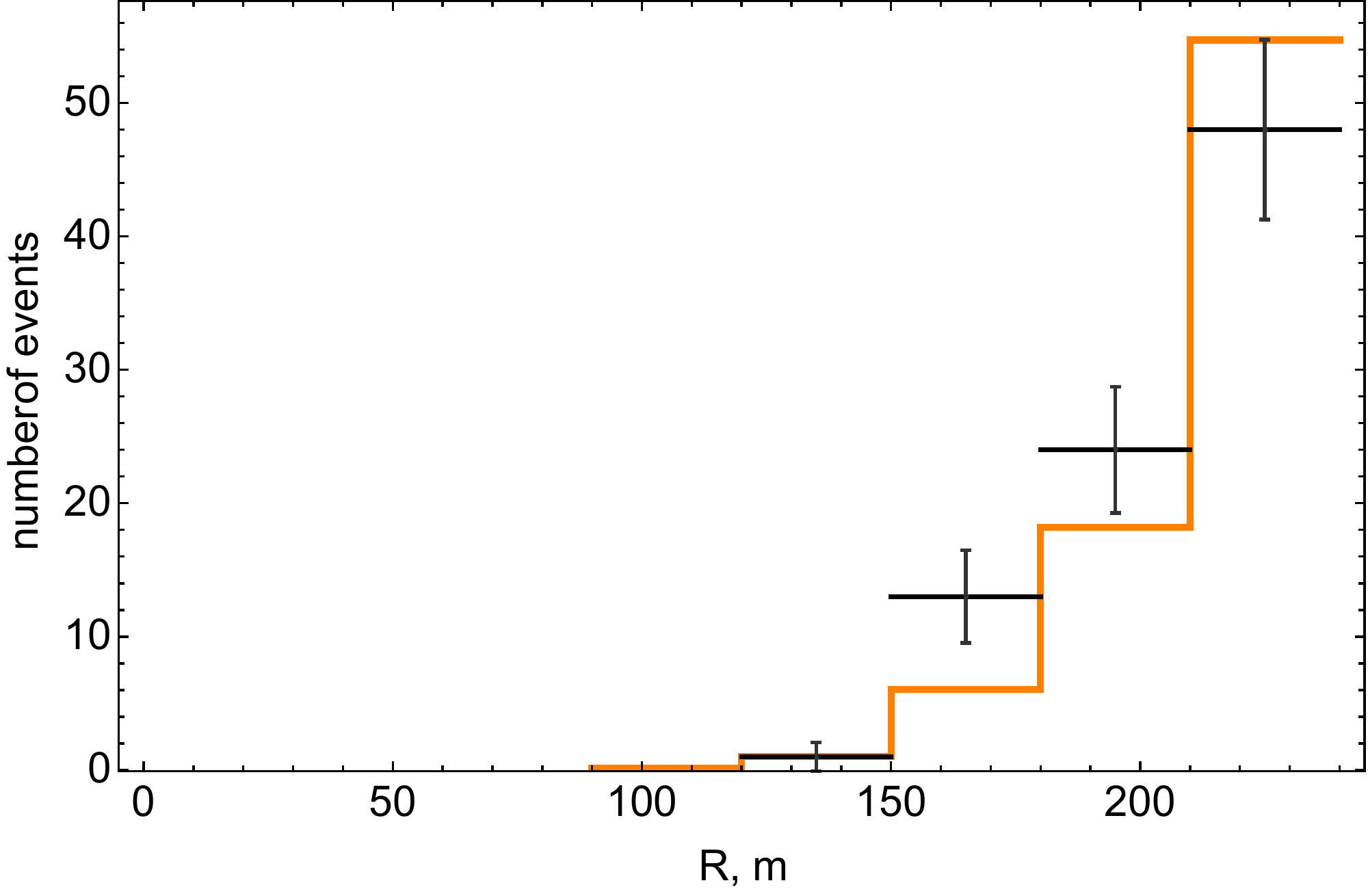} \\
\red (a) \black}
\end{minipage}
\hfill
\begin{minipage}[h]{0.48\linewidth}
\center{\includegraphics[width=1\linewidth]{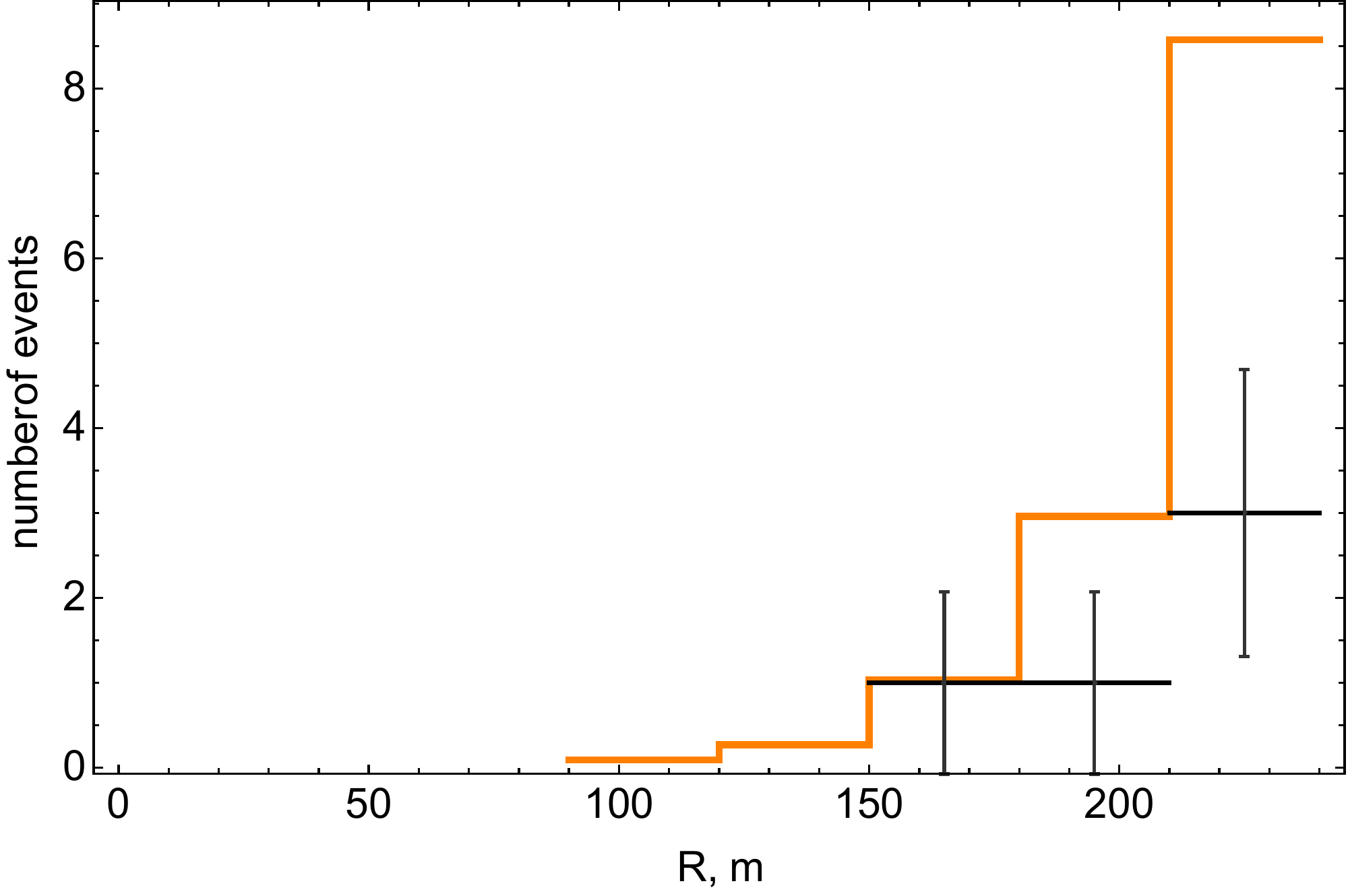}%
\\
(b)
}
\end{minipage}
\caption{Data versus MC comparison of the distribution of muonless events in $R$.
Points with error bars: data, orange hystogram: MC (mixed composition);
(a): $E_{\gamma}^{\rm min}>2\times10^{16}$~eV; (b): $E_{\gamma}^{\rm min}>10^{17}$~eV.}
\label{fig:R-comparison}
\end{figure}

The total number of muonless events in the set is 86, while the
expected number of background muonless events from primary hadrons is
80.1. The muon detector core distance distribution of the observed and
expected muonless events is shown in Fig.~\ref{fig:R-comparison}.

\begin{figure}
\centerline{\includegraphics[width=0.75\linewidth]{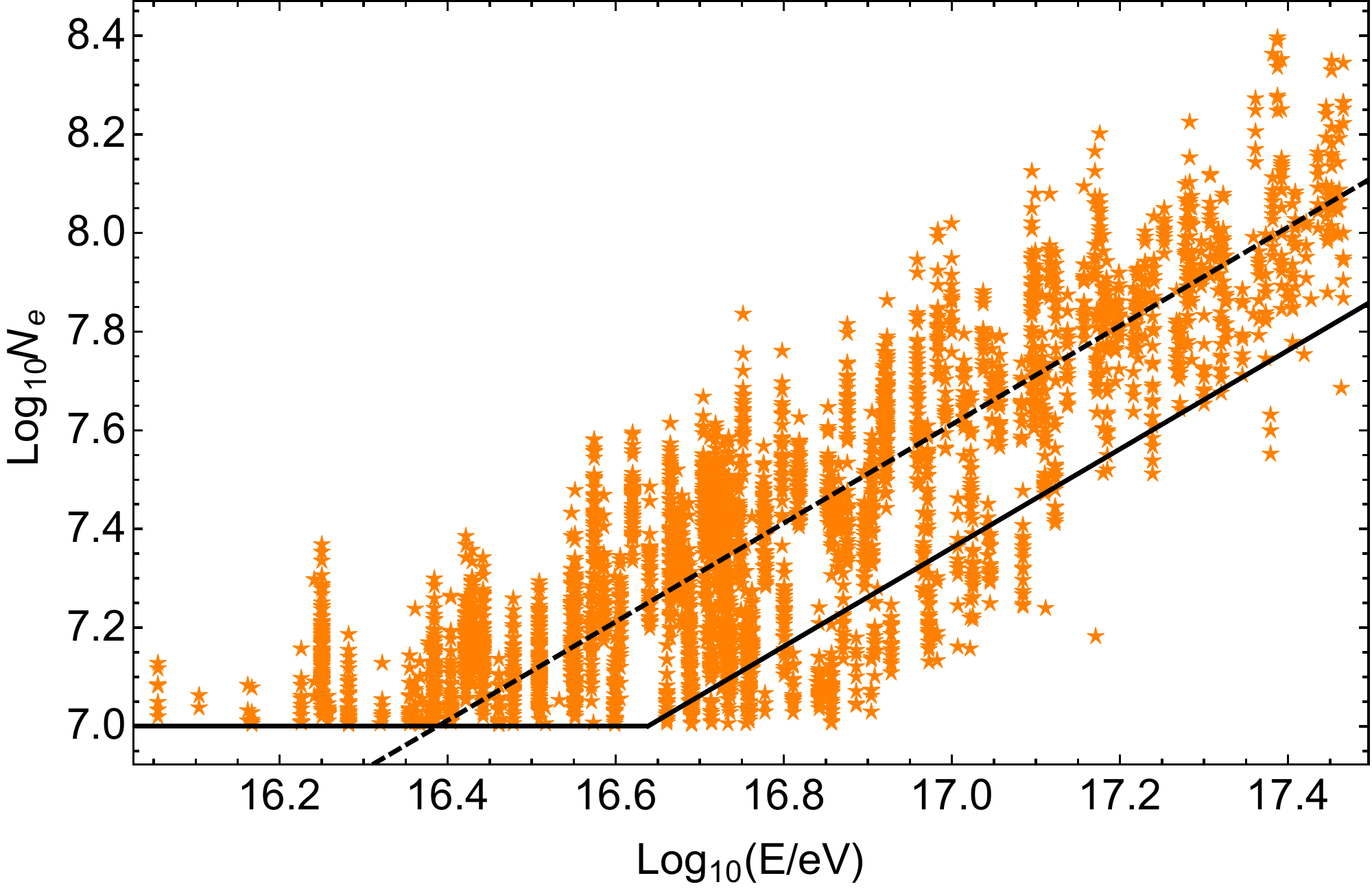}}
\caption{The $N_{e}(E_{\gamma })$ relation. Stars -- Monte-Carlo photons;
dashed line -- relation Eq.~(\ref{Eq:*}); full line bounds the region
determined by the condition Eq.~(\ref{Eq:**}). }
\label{fig:Ne-E}
\end{figure}

To study various energy ranges, we consider certain subsamples of the
data. The quantity reconstructed for each shower is $N_{e}$, not the
primary energy $E$; the relation between the two quantities may be
obtained from simulations. The $(E-N_{e})$ relations are different for
photons and for hadrons, and it is important to keep track of this
difference for gamma-ray searches, see e.g.\ Refs.~\cite{KRT-SHDM, AGYak}.
The mean gamma-ray energy $E_{\gamma}$ is related to $N_{e}$ as
\begin{equation}
N_{e}(E_{\gamma })= 4.1 \times 10^{-10} \frac{E_{\gamma }}{eV},
\label{Eq:*}
\end{equation}
see Fig.~\ref{fig:Ne-E}
for the MC simulated points and the fit.
The condition used to select the data for the search of photons with
$E_{\gamma }> E_{\gamma}^{\rm min}$ is defined as
\begin{equation}
N_{e}>\max \left\{ 10^{7}, \, a\, N_{e}(E_{\gamma}^{\rm min})   \right\},
\label{Eq:**}
\end{equation}
where the coefficient $a=0.56$ was chosen in such a way that at least 90\%
of MC photon-induced showers with $E_{\gamma }>E_{\gamma}^{\rm min}$ are
reconstructed with these $N_{e}$.

For each energy cut $E_{\gamma}^{\rm min}$, we determine the number of
observed muonless events, $n_{\rm obs}$, in the sample, as well as the
expected number of muonless events from the background of
hadron-induced showers, $n_{\rm b}$. No excess of muonless events is
seen, and we estimate the maximum number of photon-induced
events in the sample, $n_{\gamma }^{\rm FC}$, by means of the standard
Feldman--Cousins method for the Poisson distribution \cite{FC}, for
the 90\% and 95\% confidence levels (CL). The upper limits on the
  photon flux are then estimated as a ratio of $n_{\gamma }^{\rm FC}$
  and the effective exposure of the experiment to photons of the given
  energy range. The effective exposure accounts for the fraction of
photon events lost in the reconstruction and, also, of reconstructed
photon events which are not muonless. We estimate the effective exposure as
follows.

The geometrical exposure for the conditions we use ($R \le R_{\rm
  max}=240$~m, $\theta \le \theta_{\rm max}=30^{\circ}$) is given by
$A_{\rm geom}=\Omega \times S \times T$, where
$\Omega=2\pi(1-\cos\theta_{\rm max})$ is the solid angle, $S=\pi
R_{\rm max}^{2}$ is the area and $T=14060.7$ hours is
the on-time of the installation corresponding to the data set
used. Note that the $R$ cut is defined in the plane orthogonal to the
shower axis and therefore $\Omega$ is calculated differently from the
conventional case when exposure is determined by the area of the array.

MC photon-induced showers are thrown
in a square with area $S_{\rm MC}=(280~{\rm m})^{2}=0.3136$~km$^2$ and
with zenith angles up to $\theta_{\rm MC}=35^\circ$. The corresponding MC
geometrical exposure is then $A_{\rm MC}=\Omega_{\rm MC} \times S_{\rm MC}
\times T$, where $\Omega_{\rm MC}=\pi \sin^{2}\theta_{\rm MC}$ and
$\theta_{\rm MC}=35^{\circ}$. We calculate the number $n_{\rm pass, 0\mu}$
of events from the MC set which passed all cuts (that is, in particular,
were reconstructed with geometrical properties corresponding to $A_{\rm
geom}$), satisfy the criterion~(\ref{Eq:**}) and are muonless and divide it
by the number $n_{\rm MC}$ of thrown MC events (corresponding to $A_{\rm
MC}$). The effective exposure is then given by
\begin{equation}
A_{\rm eff}=\frac{n_{\rm pass, 0\mu}}{n_{\rm MC}}\,A_{\rm MC}\,.
\end{equation}
 The flux limit is then obtained as
\[
I_{\gamma }=n_{\gamma }^{\rm FC}/A_{\rm eff},
\]
where $n_{\gamma }^{\rm FC}$ corresponds to the required confidence level.
Next, we define the exposure correction as a ratio of the effective
exposure to geometrical one:
\begin{equation}
\xi = A_{\rm eff}/A_{\rm geom}\,.
\label{Eq:efficiency-coeff}
\end{equation}

\begin{figure}
\centerline{
\includegraphics[width=0.75\linewidth]{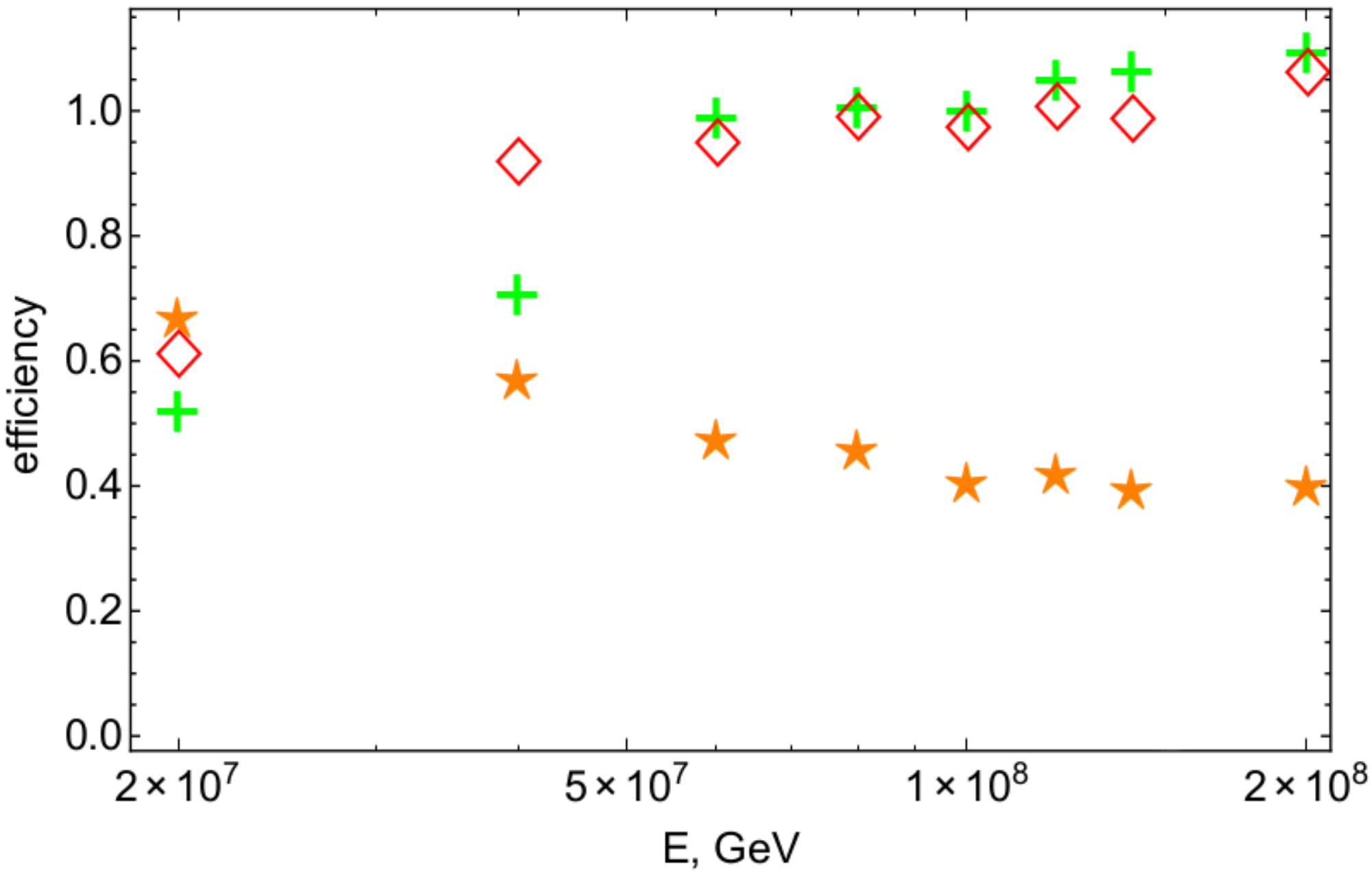}}
\caption{The exposure correction factor $\xi$ (stars) and the
reconstruction efficiency for primary protons (pluses) and photons
(diamonds) versus energy. See the text for more details.
}
\label{fig:eff}
\end{figure}

Note that the exposure correction factor may
exceed unity because Monte-Carlo events are thrown to the area larger
than the installation.
Figure~\ref{fig:eff}
presents the exposure correction factor $\xi$ as a function of energy. For
comparison, the reconstruction efficiency for primary photons and protons
is also shown (it is determined in a similar way as $\xi$ but without the
muonlessness condition and with the criterion $N_{e}>10^{7}$ instead of
Eq.~(\ref{Eq:**})). Decline of $\xi$ at higher energies reflects the
fact that the probability for a primary photon to produce a EAS which is
not muonless grows with energy.

Our limits on the integral gamma-ray flux, which represent the main result
of this work, are presented in Table~\ref{tab:flux}.
\begin{table}
\centering
\begin{tabular}{ccccccccc}
\hline
\hline
$E_{\gamma}^{\rm min}$, & $N_{e}^{\rm min}$, & $n_{\rm obs}$ & $n_{\rm b}$ & \multicolumn{2}{c}{$n_{\gamma }^{\rm FC}$} & $10^{16}\times A_{\rm eff}$,           &    \multicolumn{2}{c}{$I_{\gamma}\times10^{-16}$,~\footnotesize(s$\cdot$cm$^{2}\cdot$sr)$^{-1}$}\\
eV                      &  $10^{7}$          &               &             & \footnotesize (90\% CL)&\footnotesize(95\% CL)  &\footnotesize(s$\cdot$m$^{2}\cdot$sr) &  \footnotesize(90\% CL)          & \footnotesize(95\% CL)   \\
\hline
$ 2\times 10^{16}$       & $1$               & 86            & 80.1        & 22.4 &  25.71    &5.16                    & 4.34            & 4.98 \\
$ 4\times 10^{16}$       & $1$               & 86            & 80.1        & 22.4    &  25.71    &4.39                    & 5.09            & 5.85 \\
$ 6\times 10^{16}$       & $1.38$            & 29            & 42.6        & 2.48    &  3.85      & 3.67                 & 0.68            & 1.05 \\
$ 8\times 10^{16}$       & $1.84$            & 9             & 21.7        & 1.26 &  2.13     & 3.53               & 0.36         & 0.6  \\
$          10^{17}$      & $2.3$             & 5             & 12.9        & 1.21    & 1.84     & 3.14                   & 0.39            & 0.58\\
$ 1.2\times 10^{17}$     & $2.76$            & 4             & 8.6
& 1.66                   & 2.38                 & 3.23
& 0.51                           & 0.74\\
$ 1.4\times 10^{17}$     & $3.22$            & 2             & 5.6         & 1.44    &  2.16     & 3.05                  & 0.47               & 0.71\\
$ 2\times 10^{17}$       & $4.6$             & 1             & 2.8         & 2       &  2.75     & 3.08                  & 0.65            & 0.89\\
\hline
\hline
\end{tabular}
\caption{
\label{tab:flux}
Upper limits on the integral diffuse gamma-ray flux
$I_{\gamma}$ at photon energies $E_{\gamma }>E_{\gamma}^{\rm min}$.
The value of $N_{e}^{\rm min}$ is determined by Eq.~(\ref{Eq:**}).
$n_{\rm obs}$ is the number of observed muonless events with  $N_{e} >
N_{e}^{\rm min}$; $n_{\rm b}$ is the expected number of background
muonless events; $n_{\gamma }^{\rm FC}$ is the statistical upper limit on
the excess of muonless events over the background;
$A_{\rm eff}$ is the effective exposure for photons. $n_{\gamma }^{\rm
FC}$ and $I_{\gamma }$ are reported for two confidence levels, 90\% and
95\%, as indicated. }
\end{table}

\section{Discussion}
\label{sec:disc}
\subsection{Systematic uncertainties}
\label{sec:syst}
\begin{figure}
\centerline{
\includegraphics[width=0.75\linewidth]{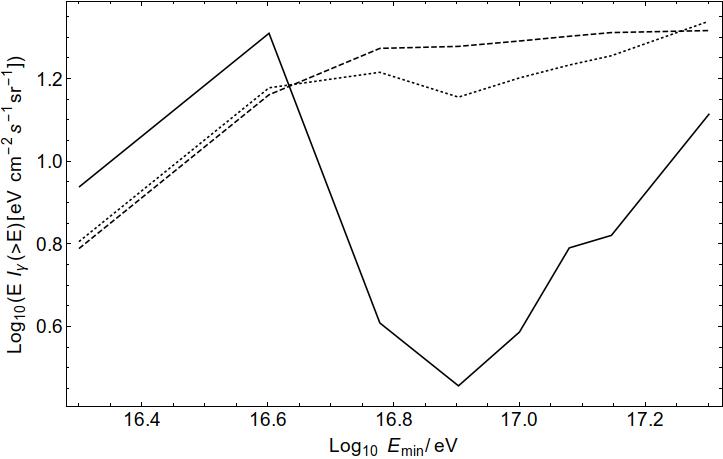}}
\caption{
Upper limits (90\% CL) on the integral flux of gamma rays under various
assumptions on the background: the MC background (full line); the
data-driven background (dotted line); the expected exclusion (dashed line).
}
\label{fig:var-bg}
\end{figure}

The main part of systematic uncertainties in the study
comes from the simulation of the background of muonless events
from hadronic showers. Indeed, it is known that hadronic-interaction
models used in the air-shower simulations are not perfect, in
particular in the part related to the description of the muon content
of EAS. While our previous study indicates~\cite{EAS-MSU_mu} that the
bulk of $E>10$~GeV muon data of the EAS-MSU experiment is well
described by the QGSJET-II-04 simulations, assuming the primary
chemical composition implied by the surface-detector studies, this
is not directly tested for muon-poor showers. The uncertainties of
the hadronic model, in principle, may reveal itself in
the incorrect estimation of the background. Indeed, we note
(cf.\ Table~\ref{tab:flux}) that for certain energy ranges, the number
of observed muonless events in our sample is smaller than expected
under the background-only hypothesis. Therefore, in addition to the
standard statistical estimate of the upper limit on the gamma-ray
flux, we calculated also the ``expected'' flux upper limits which
would be obtained if the number of observed muonless events followed
the simulations under the background-only hypothesis. Alternatively,
one may estimate the flux limits which might be obtained under
assumption that the MC model doesn't provide a reliable prediction of the
background. To this end, we use the ``data-driven background'', that
relies on the assumption that the correct background is equal to the number of the
observed muonless events. These ``expected'' and
``data-driven-background'' limits on the gamma-ray flux are presented
in Table~\ref{tab:syst} and compared in Fig.\ref{fig:var-bg}.

We also estimate the systematic errors associated with the uncertainty
of the chemical composition. The change of the proton fraction within
it's error $\pm 6\%$ results in the energy dependent correction of flux
limits. The variation of limits is 21\% for the minimum energy and 4\%
for maximum energy.
\begin{table}
\centering
\begin{tabular}{ccccc}
\hline
\hline
$E_{\gamma}^{\rm min}$, eV & \multicolumn{3}{c}{$I_{\gamma}\times
10^{16}$, (s$\cdot$cm$^{2}\cdot$sr)$^{-1}$}\\
&expected &main & data-driven \\
\hline
$ 2\times 10^{16}$       & 3.08 & 4.34 & 3.20  \\
$ 4\times 10^{16}$       & 3.62 & 5.09 & 3.76  \\
$ 6\times 10^{15}$       & 3.12 & 0.68 & 2.73  \\
$ 8\times 10^{16}$       & 2.37 & 0.36 & 1.79  \\
$          10^{17}$      & 1.95 & 0.39 & 1.59  \\
$ 1.2\times 10^{17}$     & 1.67 & 0.51 & 1.42  \\
$ 1.4\times 10^{17}$     & 1.46 & 0.47 & 1.28  \\
$ 2\times 10^{17}$       & 1.04 & 0.65 & 1.09  \\
\hline
\hline
\end{tabular}
\caption{
\label{tab:syst}
Estimate of systematic uncertainties in the flux limits: the expected
limits, the limits based on the data-driven background in comparison
with the main result of the work (90\% CL). }
\end{table}

A careful look at the relation between $n_{\rm b}$ and $n_{\rm obs}$
reveals one peculiarity which is most probably related to the modelling of
hadronic interactions but, in principle, might also be explained in terms
of the presence of a certain amount of primary photons. The ratio $n_{\rm
obs}/n_{\rm b}$ remains constant, $\sim 0.4$, for energies $E_{\gamma}
\gtrsim 8 \times 10^{16}$~eV, but quickly raises to $\sim 1$ below this
energy. If, due to some systematics in the modelling of the background,
the real $n_{\rm b}$ is indeed $\sim 0.4$ of the MC one, then we have a
certain excess of muonless events at $2\times 10^{16}$~eV$\lesssim E
\lesssim 8 \times 10^{16}$~eV, which may correspond to an excess
of primary photons expected, for instance, in the heavy dark-matter
decay scenario.

\subsection{Comparison with previous EAS-MSU results}
\label{sec:comp-EAS-MSU}
Previous analyses of the EAS-MSU data in the same energy range suggested
an excess of muonless events which might be explained by the presence of a
certain amount of photons in the primary cosmic-ray flux at $E\sim
10^{17}$~eV \cite{Khorkhe, EAS-MSU_gamma_2013, EAS-MSU_gamma_2014}. The
present study does not confirm that claim and puts strong upper limits on
the gamma-ray flux, see Table~\ref{tab:flux}. It is therefore important to
understand the differences between this analysis and the previous ones.

With respect to the previous preliminary analysis, this final study
has several important advantages. First, it is based on the new
full Monte-Carlo description of the air showers and of the
installation~\cite{EAS-MSU_Full_MC}. This uses modern simulation tools
and an analysis technique with the real and simulated events processed
by one and the same reconstruction program. In such a way, we take
into account all possible biases introduced at the reconstruction
step, as well as keep track of rare fluctuations in the EAS
development and registration. Second, the reconstruction program
  has been slightly revised for this study. The main overall effect
of the reconstruction update is that, while the muonless events
remain in the data sample, their reconstructed energies become, for
most of them, lower than before. For these lower energies, the
background of muonless hadron-induced showers is higher, and the same
amount of muonless events does no longer represent an
excess. Third, an additional check of the quality of muon data
  was performed: at least 28 out of 32 sections of the muon detector
  are required to be operational.

\begin{figure}
\centerline{\includegraphics[width=0.75\linewidth]{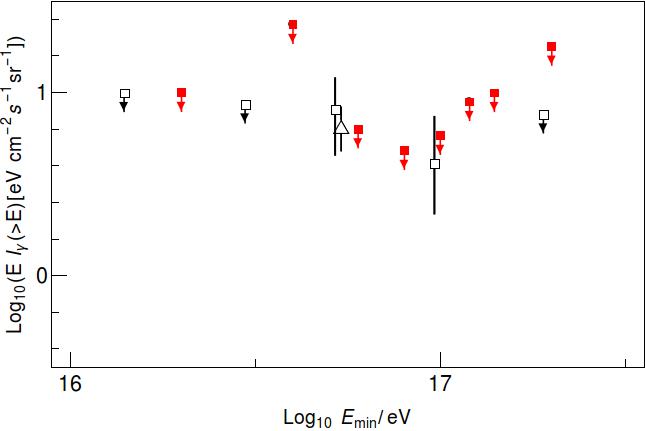}}
\caption{
Comparison of the present upper limits (95\% CL) on the integral diffuse
gamma-ray flux with previous EAS-MSU results. Full boxes (red online):
this work; open boxes: Ref.~\cite{EAS-MSU_gamma_2014}; open triangle:
Ref.~\cite{EAS-MSU_gamma_2013}.
}
\label{fig:old-vs-new}
\end{figure}

 To be
specific, let us consider 48 muonless events with $N_{e}\ge 2\times 10^{7}$
studied in Ref.~\cite{EAS-MSU_gamma_2013}. Of them, 28 events have
$N_{e}< 2\times 10^{7}$ in the new analysis; 4 events arrived before
1984 (not included in the present data set); 10 events arrived at the
days
excluded from the present analysis because of stricter criteria on the
quality of muon data;
6 muonless events remained in the
data set. In addition, 3 new muonless events joined the data set in the
new reconstruction (they did not pass the cuts in the old one), so the
total number of muonless events with $N_{e} \ge 2\times 10^{7}$ is now 9.
This reduced number of photon candidates is below the MC background
expectation of 18.9 events, so no excess is present in the updated data
set. Our new results are compared to previous studies in
Fig.\ref{fig:old-vs-new}.

Finally, the efficiency of the muonless detection of photons is
estimated in the present work with the full photon Monte-Carlo. The
numerical values of limits became apparently weaker due to account of
the fact that only about 40\% of the photon-induced showers are
registered as muonless.

\subsection{Comparison with other results and possible applications}
\label{sec:comp-other}
\begin{figure}
\centerline{\includegraphics[width=0.75\linewidth]{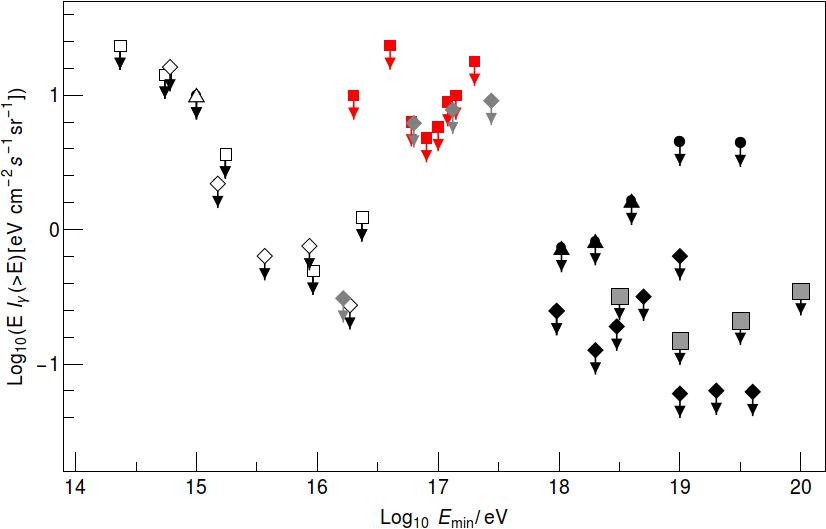}}
\caption{Limits on the integral gamma-ray flux from PeV to ZeV. Full
boxes (red online): this work;
open triangles (EAS-TOP~\cite{EAS-TOP}), open boxes
(CASA-MIA~\cite{CASA-MIA}), open diamonds (KASCADE~\cite{KASCADE}),
gray diamonds (KASCADE-Grande~\cite{KASCADE-Grande},
full triangles (Yakutsk~\cite{Yak2}), full diamonds (Pierre Auger
\cite{PAO1, PAO2}), full circles (AGASA~\cite{AGASA}), large full boxes
(Telescope Array~\cite{TA}). All limits below $10^{18}$~eV are 90\% CL,
all limits above $10^{18}$~eV are 95\% CL. }
\label{fig:all_limits}
\end{figure}

Many experiments searched for primary photons with the EAS technique and
none has yet found any. Our flux limits are compared to others in
Fig.~\ref{fig:all_limits}.
We see that our limits are similar to those of
the KASCADE-Grande experiment, being world-best for certain energies. At
the same time, one can see that the energy range $(10^{16}-10^{18})$~eV
discussed here is one of the least studied bands. While future studies to
improve the limits are important, already the present data may be used to
constrain astrophysical models with Galactic sources of PeV neutrinos \cite{Kalashev-ST-gamma}
 or decaying dark matter with appropriate mass and sufficiently hard spectrum (see \cite{Kuznetsov} for a review).

\section{Conclusions}
\label{sec:concl}
This work presents the results of the search for primary photons in the
EAS-MSU data. The photon candidate events are defined as ones
  giving
 no signal in the muon detector
of the installation. We made use of the full Monte-Carlo simulation of the
installation and of the updated reconstruction of EAS parameters. Contrary
to the previous analysis of the same data, no evidence was found
for an excess of photon-candidate events, and this fact allowed us to put
upper limits on the diffuse flux of primary gamma rays at energies
above $\sim (10^{16}-10^{17})$~eV. For certain energies, the limits are
world-best ones. The difference with the
previous study is, mainly, due to the change of reconstruction: the
energies of muonless events moved downwards, while the background of
muonless showers from primary hadrons is higher at lower energies. The
limits obtained in this work may be used in multimessenger astrophysics as
well as for constraining exotic particle-physics models.

\section*{Acknowledgements}
ST thanks J.-M.~Fr\`ere for a discussion on the data-driven background and
the Service de
Physique Th\'{e}orique at Universit\'{e} Libre de Bruxelles for
hospitality at the final stages of the work. Monte-Carlo simulations have
been performed at the computer cluster of the Theoretical Physics
Department, Institute for Nuclear Research of the Russian Academy of
Sciences. The experimental work of the EAS-MSU group is supported by the
Government of the Russian Federation (agreement 14.B25.31.0010) and the
Russian Foundation for Basic Research (project 14-02-00372). Development
of the analysis methods and application of them to the EAS-MSU data is
supported by the Russian Science Foundation (grant 14-12-01340).

\section*{Appendix. Distribution of EAS parameters for data and Monte-Carlo}

\begin{figure}[!h]
\centerline{\includegraphics[width=0.75\linewidth]{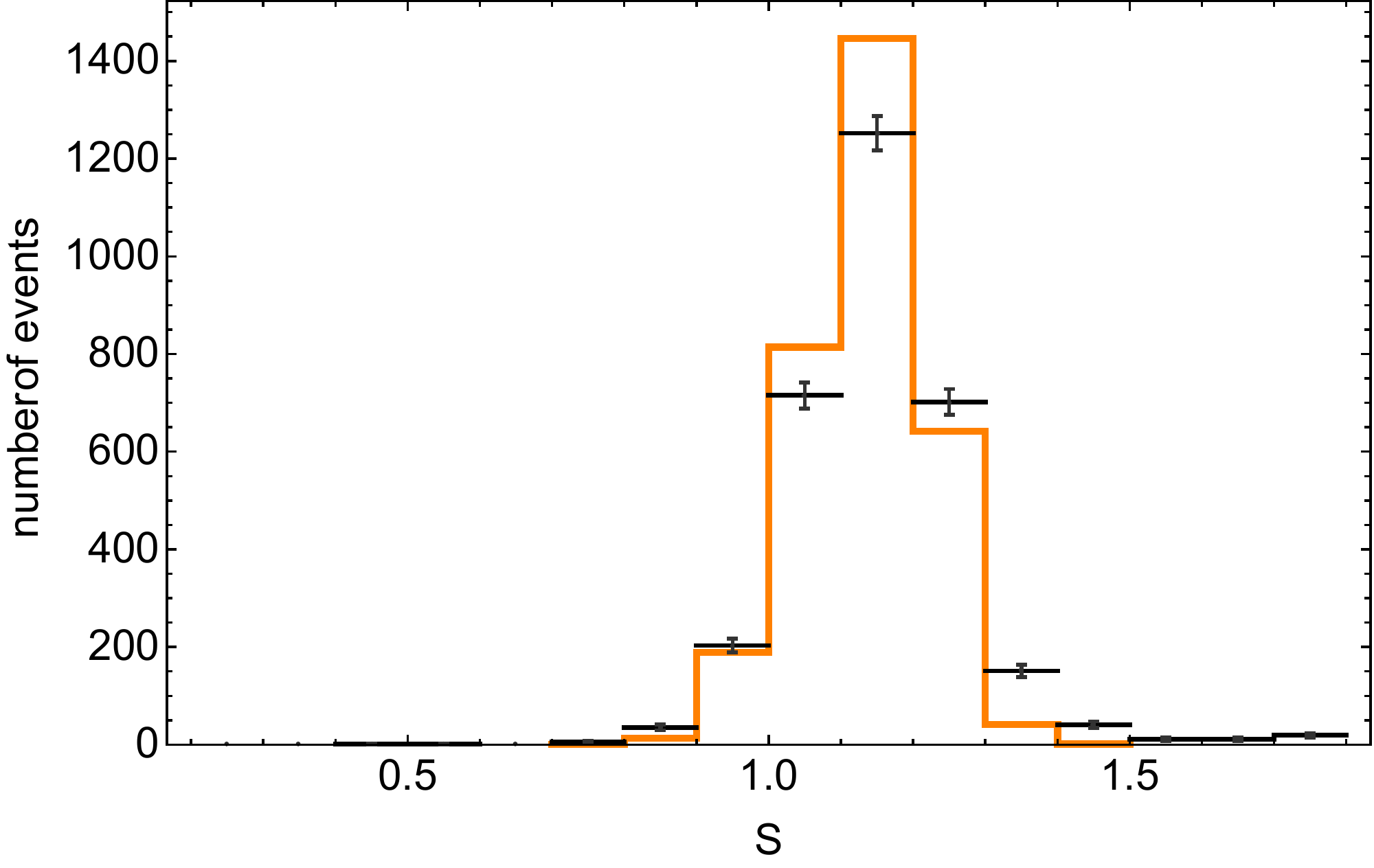}}
\caption{
$S$ distribution for data (points) and MC (orange line).
\label{S_dis}}
\end{figure}

\begin{figure}[!h]
\centerline{\includegraphics[width=0.75\linewidth]{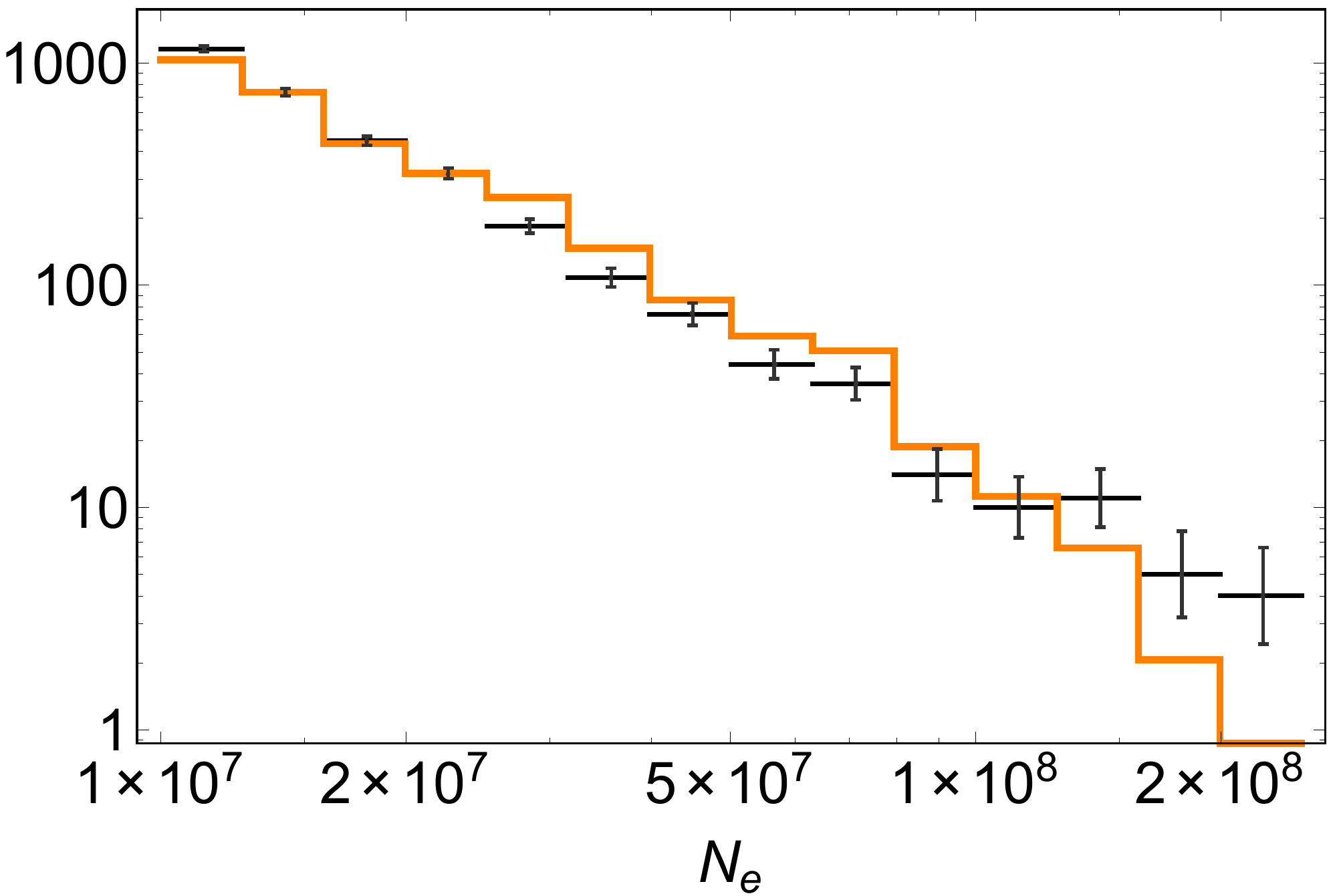}}
\caption{
$N_{e}$ distribution for data (points) and MC (orange line).
\label{Ne_dis}}
\end{figure}

\begin{figure}[!h]
\centerline{\includegraphics[width=0.75\linewidth]{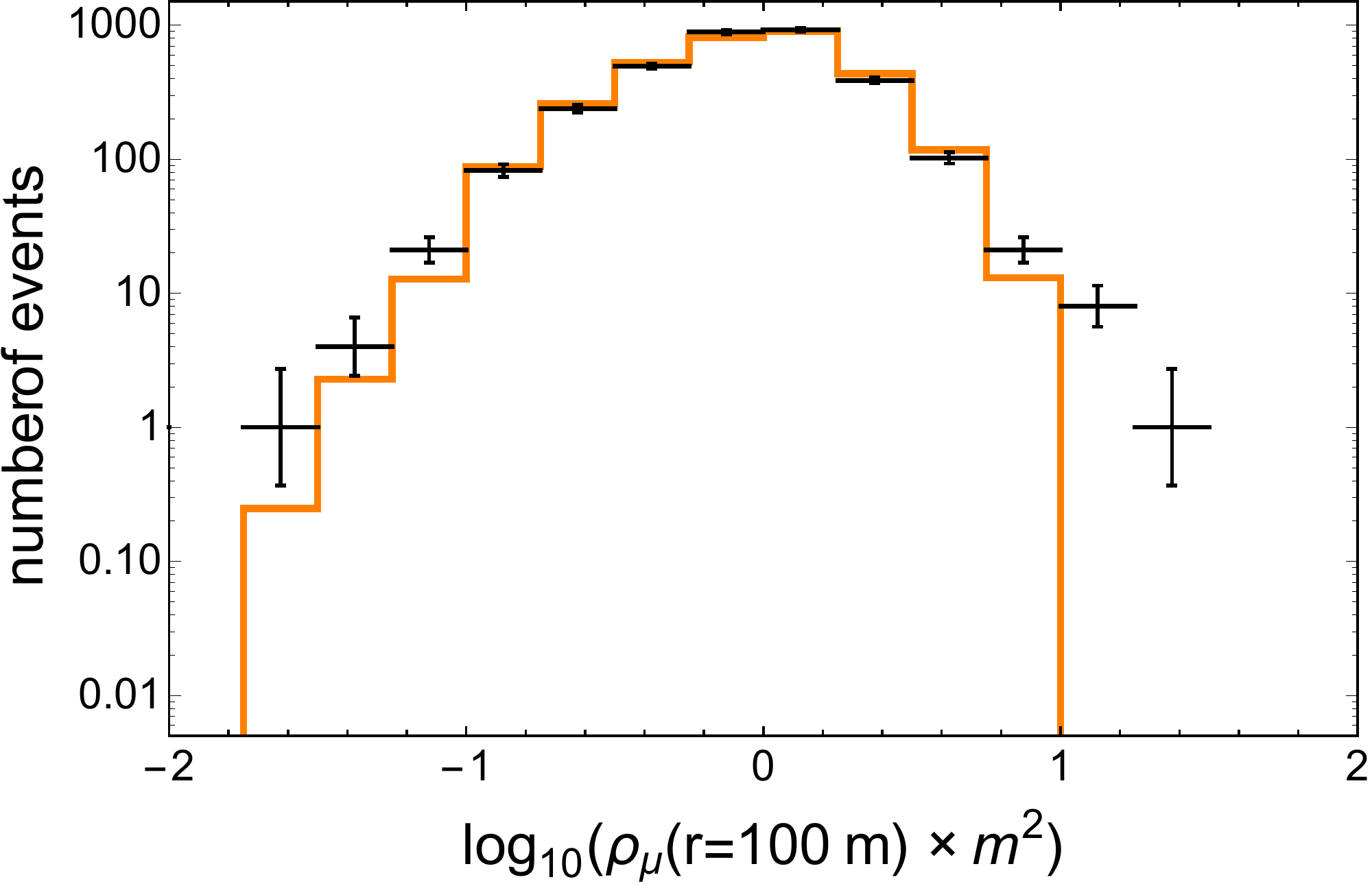}}
\caption{
$\rho_{\mu}(100)$ distribution for data (points) and MC (orange line).
\label{Ro_mu_dis}}
\end{figure}

\begin{figure}[!h]
\centerline{\includegraphics[width=0.75\linewidth]{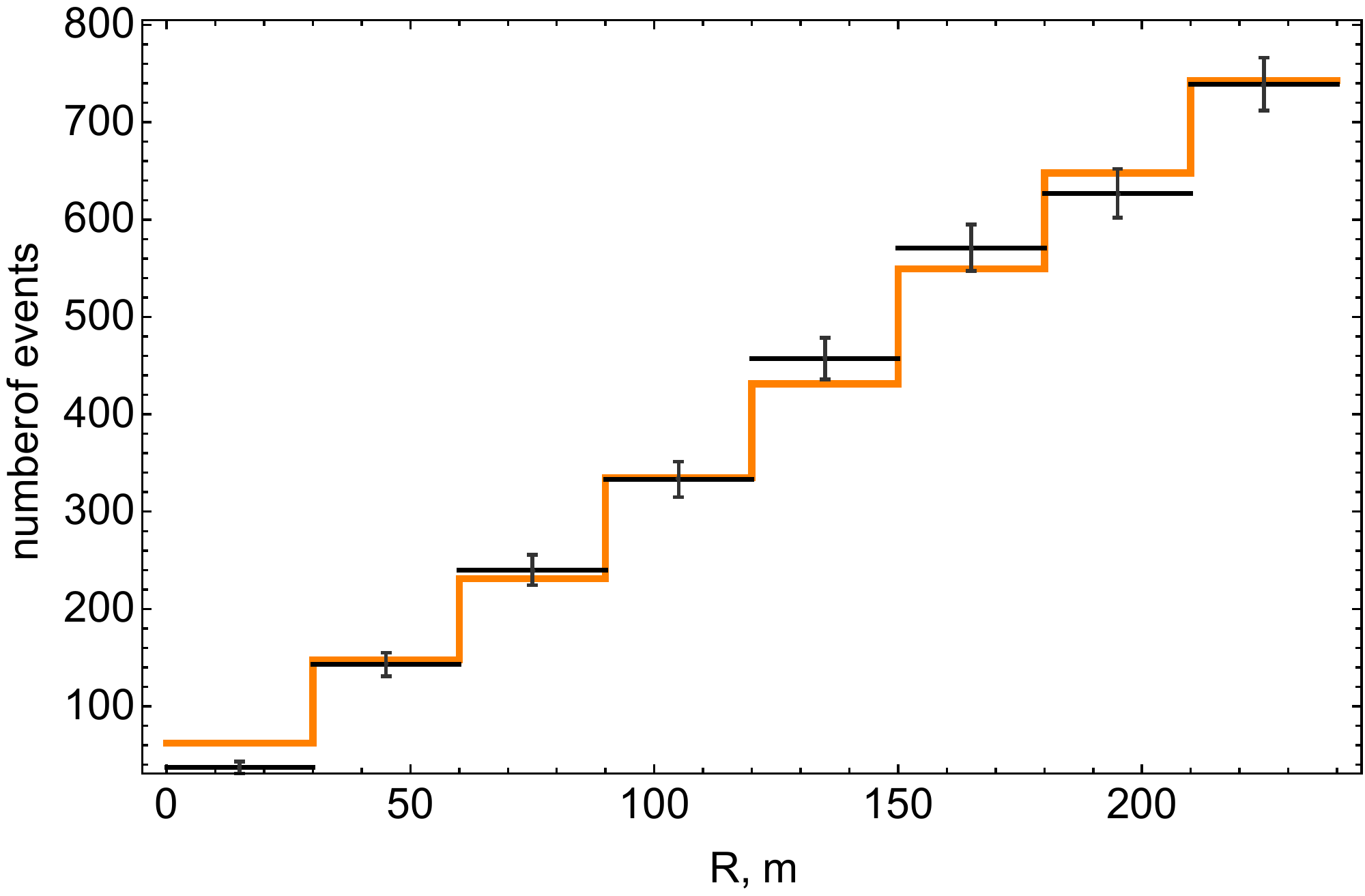}}
\caption{
$R$ distribution for data (points) and MC (orange line).
\label{R_dis}}
\end{figure}

\begin{figure}[!h]
\centerline{\includegraphics[width=0.75\linewidth]{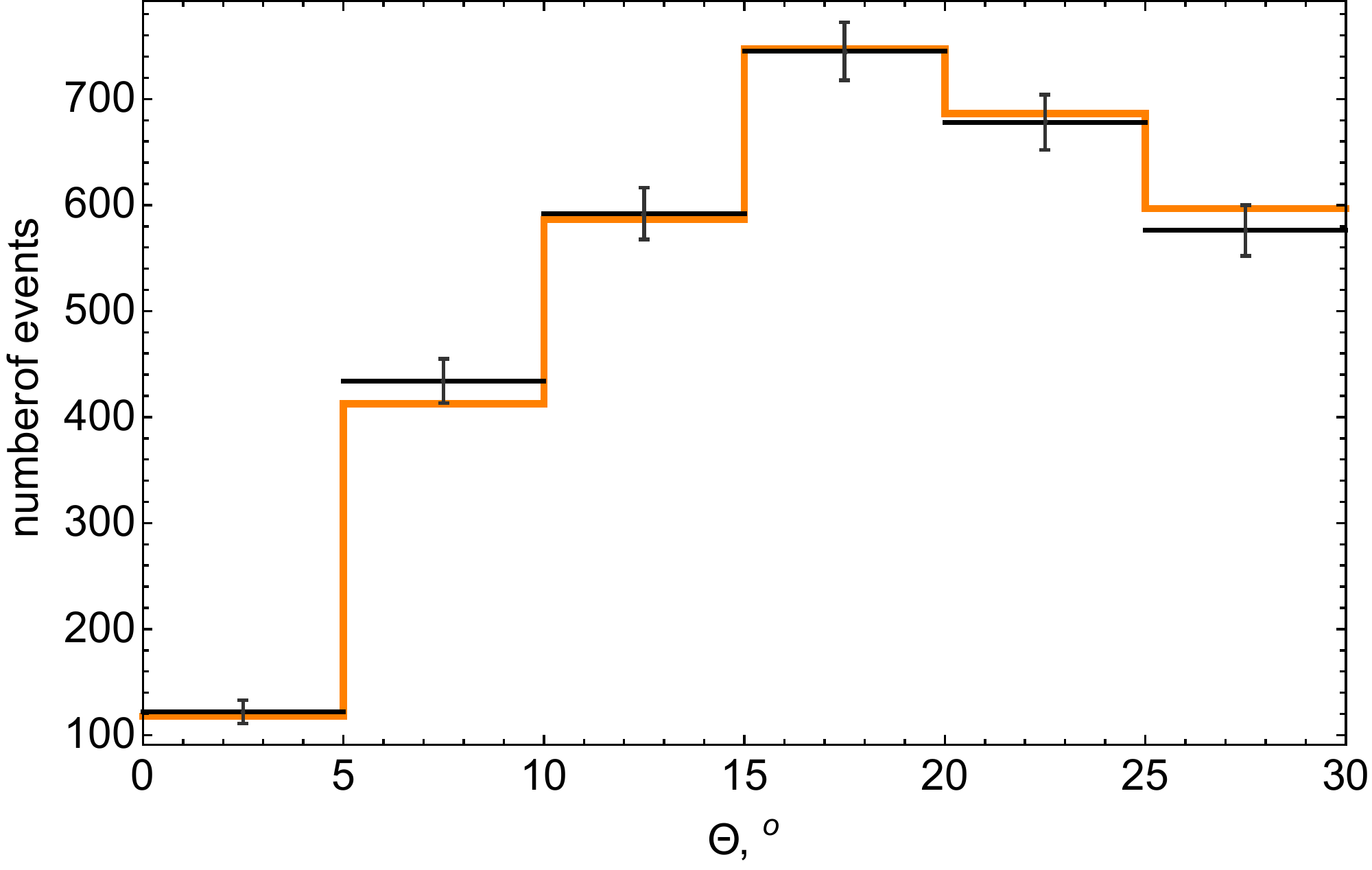}}
\caption{
$\theta$ distribution for data (points) and MC (orange line).
\label{Theta_dis}}
\end{figure}

  In this appendix, we provide the data to Monte-Carlo comparison
  of the EAS parameter distribution for the extended cut
  $N_{e}>10^{7}$ used is the present work. The primary composition is
  the same as determined in Ref.~\cite{EAS-MSU_mu} by fitting the
  observed distribution of the muon densities: $46\%$ protons and
  $54\%$ iron. The distributions of $S$, $N_{e}$, $R$, $\theta$ and
  muon density at 100 meters $\rho_{\mu}(100)$ are shown in
  Figures~\ref{S_dis}-\ref{Theta_dis}.  The validity of the $N_{e}$
  cut extension is verified by reasonable agreement of the data and
  Monte-Carlo.

\end{document}